\documentclass[aps,prl,twocolumn,groupedaddress]{revtex4}
\usepackage{epsfig}
\usepackage{amsmath}

\begin{document}

\preprint{OSU/203-MgCmPRL}

\title{Measurement of the Survival Probabilities for Hot Fusion Reactions}

\author{R. Yanez}
\author{W. Loveland}
\author{L. Yao}
\author{J. S. Barrett}
\affiliation{Department of Chemistry, Oregon State University,
 Corvallis, OR, 97331.}
 \author{S. Zhu}
\author{B.B. Back}
\author{T.L. Khoo}
\author{M. Alcorta}
\thanks{Present address: TRIUMF, Vancouver, British Columbia V6T 2A3, Canada}
\author{M. Albers}
\affiliation{Physics Division, Argonne National Laboratory,
Argonne, IL 60439.}

\date{\today}

\begin{abstract}

We have studied  the fission-neutron emission competition in highly excited $^{274}$Hs (Z=108) (where the fission barrier is due to shell effects) formed by a hot fusion reaction.
Matching cross bombardments ($^{26}$Mg + $^{248}$Cm and $^{25}$Mg + $^{248}$Cm) were used to identify the properties of first chance fission of $^{274}$Hs.  A Harding-Farley analysis of the fission neutrons emitted in the $^{25,26}$Mg + $^{248}$Cm was performed  to identify the pre- and post-scission components of the neutron multiplicities in each system. ($\Gamma$$_{n}$/$\Gamma$$_{t}$) for the first chance fission of $^{274}$Hs (E$^{\ast}$ = 63 MeV) is 0.89 $\pm$ 0.13, i.e., $\sim$ 90 $\%$ of the highly excited nuclei survive.The high value of that survival probability is due to dissipative effects during de-excitation. A proper description of the survival probabilities of excited superheavy nuclei formed in hot fusion reactions requires consideration of both dynamic and static (shell-related) effects.

\end{abstract}


\maketitle

The remarkable recent progress in the synthesis of new heavy and superheavy nuclei has been made using fusion reactions.  These reactions can be divided into two prototypical classes, ``cold"  and ``hot" fusion reactions.  In ``cold" fusion reactions, one bombards Pb or Bi target nuclei with heavier projectiles (Ca-Kr) to form completely fused systems with low excitation energies ($E^\ast$=10-15 MeV), leading to a higher survival (against fission) but with a reduced probability of the fusion reaction taking place due to the larger Coulomb repulsion in the more symmetric reacting system.  (This approach has been used in the synthesis of elements 107-113).  In ``hot" fusion reactions one uses a more asymmetric reaction (typically involving a lighter projectile and an actinide target nucleus) to increase the fusion probability but leading to a highly excited completely fused system ($E^\ast$=30-60 MeV) with a reduced probability of surviving against fission.  (This approach has been used to synthesize elements 102-118.)

Formally, the cross section for producing a heavy evaporation residue, $\sigma$$_{\rm EVR}$, in a fusion reaction can be written as
\begin{equation}
\sigma _{\rm EVR}=\sum_{J=0}^{J_{\max }}\sigma
_{\rm capture}(E_{\rm c.m.},J)P_{\rm CN}(E^\ast,J) W_{\rm sur}(E^\ast,J)
\end{equation}
where $\sigma _{\rm capture}(E_{\rm c.m.},J)$ is the capture cross section at center of mass energy E$_{\rm c.m.}$ and spin J. P$_{\rm CN}$ is the probability that the projectile-target system will evolve from the contact configuration  inside the
fission saddle point to form a completely fused system rather than
re-separating (quasifission, fast fission). W$_{\rm sur}$ is
the probability that the completely fused system will de-excite by neutron emission rather than fission. For a quantitative understanding of the synthesis of new heavy nuclei, one needs to understand $\sigma _{\rm capture}$, P$_{\rm CN}$, and W$_{\rm sur}$ for the reaction system under study.
	
	Formally W$_{sur}$ can be written as 
\begin {equation}
W_{sur}(E^{\ast },J)=P_{xn}(E^{\ast })\prod\limits_{i=1}^{x}\left( \frac{%
\Gamma _{n}}{\Gamma _{n}+\Gamma _{f}}\right) _{i,E^{\ast },J}
\end{equation}
where P$_{xn}$ is the probability of emitting x (and only x) neutrons from a nucleus with excitation energy E$^{\ast}$, and $\Gamma$$_{n}$ and $\Gamma$$_{f}$ are the partial widths for decay of the completely fused system by either neutron emission or fission, respectively.  (Decay of the completely fused system by charged particle emission is neglected)

	The survival probability $W_{\rm sur}(E^\ast,J)$ is of intrinsic interest.  For example, the reported cross sections for the synthesis of superheavy elements in hot fusion reactions decrease modestly in going from element 113 to element 118. (Figure 1) This behavior is attributed to the increasing survival probability of the product nuclei as one gets closer to Z=114 or N=184 shells. \cite{ogarev}.  The real situation is complicated with the fused systems starting at excitation energies, E*, of 30-50 MeV where shell effects on $\Gamma$$_{n}$/$\Gamma$$_{f}$ are not important but where dissipative effects \cite{kramers} may retard fission and ending at low excitation energies where shell effects are very important.

In this work, we are extending the studies of survival probabilities in hot fusion reactions  to study the $^{25,26}$Mg + $^{248}$Cm reaction which leads to the formation of Hs (Z=108) isotopes.  The evaporation residue yields in this reaction have been studied before \cite{tur,dv1,dv2} and cross sections are available for the formation of $^{269,270,271}$Hs.  These evaporation residues are especially interesting because they are at or near the proposed deformed shell at N=162, Z=108 \cite{patyk}.  As such, they form a perfect laboratory for the study of shell effects on the survival of highly excited nuclei.  P$_{CN}$ is known to be 1.0 (no quasifission) for this system \cite{itkis,koz,rycn} leading to an unambiguous relation between the observed neutron multiplicity and W$_{sur}$.

There is considerable controversy regarding the appropriate values of W$_{sur}$ and/or the relevant fission barrier heights, B$_{f}$,  for the $^{248}$Cm($^{26}$Mg, xn) systems.   The Extended Thomas Fermi Strutinski Integral (ETFSI) estimated barrier height for $^{274}$Hs is 2.50 MeV \cite{ETFSI}, the Finite Range Liquid Drop Model barrier height was 5.31 MeV \cite{peter1} but has been recently revised to be 6.45 MeV \cite{peter2}, the macroscopic-microscopic locally tuned barrier height is 4.37 MeV \cite {kowal}, while a recent Density Functional Theory calculation \cite{pg} gave a barrier height for $^{274}$Hs of 5.1 $\pm$ 0.5 MeV.  

There has been one previous attempt to characterize the fission neutrons emitted in the 160 MeV $^{26}$Mg + $^{248}$Cm reaction \cite{itkis1,itkis2}.  The total fission neutron multiplicity was 12 $\pm$1, with the pre-scission neutron multiplicity  $\nu_{pre}$ = 4 $\pm$ 1 and the post-scission neutron multiplicity  $\nu_{post}$ = 4.5 $\pm$1.  These numbers refer to the entire evaporation chain (7 chances to fission) and are somewhat surprising.  Standard fission theory \cite{nrvp} would predict $\nu_{pre}$ = 2.7 and $\nu_{post}$ = 7.5.  Our measurement will also measure these quantities, but more importantly, we will isolate the contribution of first chance fission.

   The methods are to be the same as used on the study of the de-excitation of $^{258}$No, \cite{don} i.e., to form $^{274}$Hs (E*= 63 MeV) via the $^{26}$Mg + $^{248}$Cm reaction and to form $^{273}$Hs(E*= 53 MeV) using the $^{25}$Mg + $^{248}$Cm reaction and to measure the fission associated neutron multiplicities in these reactions. The logic behind the use of cross bombardments is given below.

In the reaction $^{26}$Mg~+~$^{248}$Cm, we detected
neutrons from the decay chain 
\begin{equation}
^{274}Hs \rightarrow ^{273}Hs \rightarrow ^{272}Hs \rightarrow ^{271}Hs \rightarrow ^{270}Hs \rightarrow\cdots \\
\end{equation}
For the $^{25}$Mg + $^{248}$Cm reaction, the bombarding energy was chosen
such that its decay chain 
\begin{equation}
{^{273}Hs \rightarrow ^{272}Hs \rightarrow^{271}Hs \rightarrow ^{270}Hs \rightarrow \cdots }
\end{equation}
energetically matches the corresponding elements of the $^{274}$Hs chain,
allowing a direct comparison of the neutrons emitted solely from that
nucleus avoiding the difficulties inherent to other, model-dependent
analysis methods.  The issue of whether the technique of cross bombardments will actually allow one to 
deduce the properties of the first members of the de-excitation chains since we expect 
differences in the spin distributions of the products formed in these reactions was treated previously \cite{don} where it was demonstrated that the fission-neutron emission competition in the two reactions is expected to be similar.

For the reaction, $^{26}$Mg + $^{248}$Cm producing $^{274}$Hs, the number of pre-scission neutrons, $\nu_{pre}$ is given as 
\begin{equation}
\nu _{pre}^{274}\geq \left( \frac{\Gamma _{n}}{\Gamma _{tot}}\right)
_{1}+\left( \frac{\Gamma _{n}}{\Gamma _{tot}}\right) _{1}\left( \frac{\Gamma_{n}}{\Gamma _{tot}}\right) _{2}+\cdots 
\end{equation}
while for the $^{25}$Mg + $^{248}$Cm reaction producing $^{273}$Hs, we have
\begin{equation}
\nu _{pre}^{273}\geq \left( \frac{\Gamma _{n}}{\Gamma _{tot}}\right)
_{2}+\left( \frac{\Gamma _{n}}{\Gamma _{tot}}\right) _{2}\left( \frac{\Gamma_{n}}{\Gamma _{tot}}\right) _{3}+\cdots 
\end{equation}
where the subscripts refer to the n$^{th}$ chance to fission.  One can show that 
\begin{equation}
\left( \frac{\Gamma _{n}}{\Gamma _{tot}}\right) _{1}=\frac{\nu
_{pre}^{274}}{1+\nu _{pre}^{273}}
\end{equation}
Thus measuring $\nu_{pre}$ for each reaction will allow one to deduce  $\left( \frac{\Gamma _{n}}{\Gamma _{tot}}\right) $ for the first chance fission of $^{274}$Hs.

The experimental apparatus is shown in Figure 2.  A thin-walled Al scattering chamber was used to hold the target and fission detectors with an external array of BC-501 scintillators to detect the neutrons.  

The Mg beams were accelerated by the ATLAS accelerator at the Argonne National Laboratory. This resulted in center-of-target reaction energies of 160.7 MeV
and 144.2 MeV, respectively. The $^{248}$Cm target consisted of a deposit of 79.9 $\mu$g/cm$^{2}$ of $^{248}$Cm on a 0.54 mg/cm$^{2}$ Al foil.  The area of the deposit was 0.071 cm$^{2}$. The target was tilted at 45$^{\circ }$\ with respect to the beam
axis. 

Fission fragments were detected by an array of seven Si PIPS detectors (300 mm$^{2}$) positioned at 16.1 cm from the target at various angles.  Neutrons were detected in coincidence with fission by an array of six BC501A scintillators arranged in  a plane as shown in Figure 2.

Neutron time of flight (TOF) spectra were obtained from the time difference
between a signal from one of the fission detectors and a signal from a scintillator. Fragment TOFs were
obtained from the time difference between an accelerator RF signal  and a Si detector. Neutron
and $\gamma$-ray events in the detectors were separated by means of pulse
shape discrimination in the off-line analysis.  After appropriate clean-up, less than 1 $\%$ of the neutron events were due to photons. These events will correspond to high energy neutrons that will have a negligible effect on the multiplicities.   A $^{137}$Cs source was used to set the thresholds of the constant fraction discriminators for the neutron signals at $\sim$ 350 keV \cite{master}.

The efficiencies of the neutron detectors were obtained using a $^{252}
$Cf fission source placed at the target location. The neutron energy spectrum was assumed to have three components \cite{btms} represented by analytic functions of the form
\begin{equation}
\phi (n)=(\frac{n}{T_{i}^{2}})exp\left ( -n/T_{i} \right )
\end {equation}
where the T$_{i}$ (temperature) values were 0.9941, 0.3729, and 0.0731, the weighting factors of each component were 0.5720, 0.4061, and 0.0219, respectively and $\eta$ is the neutron energy in the center of mass.  The three components represent emission from the two moving fragments and isotropic emission from a source at rest in the laboratory system.  The mean fission fragment energies and masses were taken to be 80.6 and 106.2 MeV and 143.6 and 108.4, respectively \cite{stan}. 

MCNP5 \cite{mcnp5} was used to calculate the transport of the neutrons from the $^{252}$Cf source to the entrance of the BC501A detectors.  Then SCINFUL \cite{dickens} was used to associate each incoming neutron energy to a light output.    As a consistency check, one can use this efficiency function to measure the mean neutron multiplicity for the spontaneous fission of $^{252}$Cf (3.74 $\pm$ 0.05) in good agreement with the known value. Efficiency measurements were made
before and after the bombardments.

To determine the pre- and post-scission neutron multiplicities, a multi-source fitting procedure was used.  Neutrons were assumed to be emitted isotropically in the rest frames of three moving sources, the compound nucleus, and the two fully accelerated fission fragments.  
The resulting fits to the data are shown in Figure 3 for $^{274}$Hs.  The numerical values of the fitting parameters are given in Table 1.  Using equation [7], and values of $\nu _{pre}^{274}$ = 3.08 $\pm$ 0.31 and  $\nu _{pre}^{273}$ = 2.48 $\pm$ 0.25, we can calculate that $\Gamma$$_{n}$/$\Gamma$$_{total}$ for the first chance fission of $^{274}$Hs (E$^{*}$ = 60 MeV) is 0.89 $\pm$ 0.13, i.e., about 90 $\%$ of these highly excited nuclei survive fission decay.

To help understand our measured value of $\Gamma$$_{n}$/$\Gamma$$_{total}$ for the first chance fission of $^{274}$Hs, we have calculated the expected value of this ratio using the formalism of \cite{nrvp}.  We took the ground state (and saddle point) masses and deformations, and shell corrections from \cite{peter1, audi}.  We used the macroscopic-microscopic locally tuned barrier height of 4.37 MeV \cite {kowal}.  The level density parameter and its excitation energy dependence were taken from \cite{iggy} with  $\gamma$ =0.061, where
\begin{equation}
a=\widetilde{a}\left[ 1+\delta E\frac{1-\exp (-\gamma E)}{E}\right] 
\end{equation}
\begin{equation}
\widetilde{a}=0.073A+ 0.095B_{s}(\beta _{2})A^{2/3}
\end {equation}

Collective enhancement effects of the level density are important for both deformed and spherical nuclei as are their dependence on excitation energy.   We used  the formalism of \cite{zaggy, jung} to express these effects via the equations
\begin{equation}
K_{coll}=K_{rot}(E)\varphi (\beta _{2})+K_{vib}(E)\cdot (1-\varphi(\beta _{2}))
\end{equation}
\begin{equation}
\varphi (\beta _{2})=\left[ 1+\exp \left( \frac{\beta _{2}^{0}-\left\vert\beta _{2}\right\vert }{\Delta \beta _{2}}\right) \right] ^{-1}
\end{equation}
\begin{equation}
K_{rot(vib)}(E)=\frac{K_{rot(vib)}-1}{1+\left[ \left( E-E_{cr}\right)/\Delta E_{cr }\right] }+1
\end{equation}
\begin{equation}
K_{rot}=\frac{J_{\bot }T}{\hbar ^{2}}
\end{equation}
\begin{equation}
K_{vib}=\exp (0.0555A^{2/3}T^{4/3})
\end{equation}
where K$_{vib}$ is taken from \cite{iggybook}, E$_{cr}$ = 40 MeV and $\Delta$E=10 MeV.  

Calculations of the widths were done for individual $\ell$ values and weighted by the probability of populating that $\ell$ value.  The calculated value of $\Gamma$$_{n}$/$\Gamma$$_{total}$ for the first chance fission of $^{274}$Hs was 0.18  compared to the measured value of 0.89 $\pm$ 0.13.

To determine the effect of the assumed value of B$_{f}$ upon the calculation, we performed another calculation with this formalism assuming a fixed $\ell$ value of 35 $\hbar$ and varied B$_{f}$.  In Figure 4, we show the resulting values of $\Gamma$$_{n}$/$\Gamma$$_{total}$ as a function of B$_{f}$.  For the range of B$_{f}$ values predicted by current models (Section IA) , there is very little sensitivity to B$_{f}$.  To get values of  $\Gamma$$_{n}$/$\Gamma$$_{total}$ similar to the measured value, one must invoke values of B$_{f}$ $\sim$ 11 MeV, a number not consistent with current theoretical models for superheavy nuclei.

Lestone \cite{lessy} has suggested that considering the tilting degree of freedom at the saddle point will lead to a reduction of the fission probability.  Formally
\begin{equation}
\Gamma _{f}^{Lestone}=\Gamma _{f}^{BW}\frac{K_{0}\sqrt{2\pi }}{2J+1}erf\left ( \frac{J+1/2}{\sqrt{2}K_{_{0}}} \right )
\end{equation}
where $\Gamma^{BW}_{f}$ is the conventional transition state calculation of the fission width.  Making this correction in the asymptotic limit increased $\Gamma$$_{n}$/$\Gamma$$_{total}$ from 0.18 to 0.23.

Kramers \cite{kramers} pointed out that the fission width as calculated by the standard Bohr-Wheeler transition state theory, $\Gamma_{f}^{BW}$ is reduced by the effect of nuclear viscosity.  Formally, this Kramers correction can be expressed as
\begin{equation}
\Gamma _{f}^{Kramers}=\Gamma _{f}^{BW}\left ( \sqrt{1+\gamma ^{2}} -\gamma \right )
\end{equation}
where $\gamma$ is a dimensionless dissipation coefficient.  Using $\gamma$ as a free parameter, the observed value of $\Gamma$$_{n}$/$\Gamma$$_{total}$  for the first chance fission of $^{274}$Hs (E*= 63 MeV) can be reproduced if $\gamma$ $\sim$ 18.  This relatively large value of $\gamma$ is consistent with previous work on pre-scission giant dipole resonance gamma-ray emission \cite{hbp} and suggestions \cite {tb,v} of high pre-scission multiplicities in systems of high fissility.

We conclude that : (a) We have made the first measurement of the first chance survival probability of a highly excited superheavy nucleus where the fission barrier is due solely to nuclear shell effects. (b) The high value of that survival probability ($\Gamma_{n}$/$\Gamma_{total}$ = 0.89$\pm$ 0.13 ) is due to dissipative effects during de-excitation.  (c) A proper description of the survival probabilities of excited superheavy nuclei formed in hot fusion reactions requires consideration of both dynamic and static (shell-related) effects.

This work was supported in part by the
Director, Office of Energy Research, Division of Nuclear 
Physics of the Office of High Energy and Nuclear Physics 
of the U.S. Department of Energy
under Grant DE-FG06-97ER41026 and contract No. DE-AC02-06CH11357.

\begin{table}
\caption{Values of fitting parameters}
\label{tab:1}       
\begin{tabular}{ccccc}
\hline\noalign{\smallskip}
Compound Nucleus& T$^{pre}$& T$^{post}$ & M$^{pre}$ & M$^{post}$  \\
\noalign{\smallskip}\hline\noalign{\smallskip}
$^{274}$Hs&0.38$\pm$0.02&0.72$\pm$0.02& 3.08$\pm$0.31&8.56$\pm$0.02 \\
$^{273}$Hs&0.36$\pm$0.02&0.67$\pm$0.02&2.48$\pm$0.25&7.72$\pm$0.04\\
\noalign{\smallskip}\hline
\end{tabular}
\end{table}

\begin{figure}[tbp]
\begin{minipage}{30pc}
\begin{center}
\includegraphics [width=150mm]{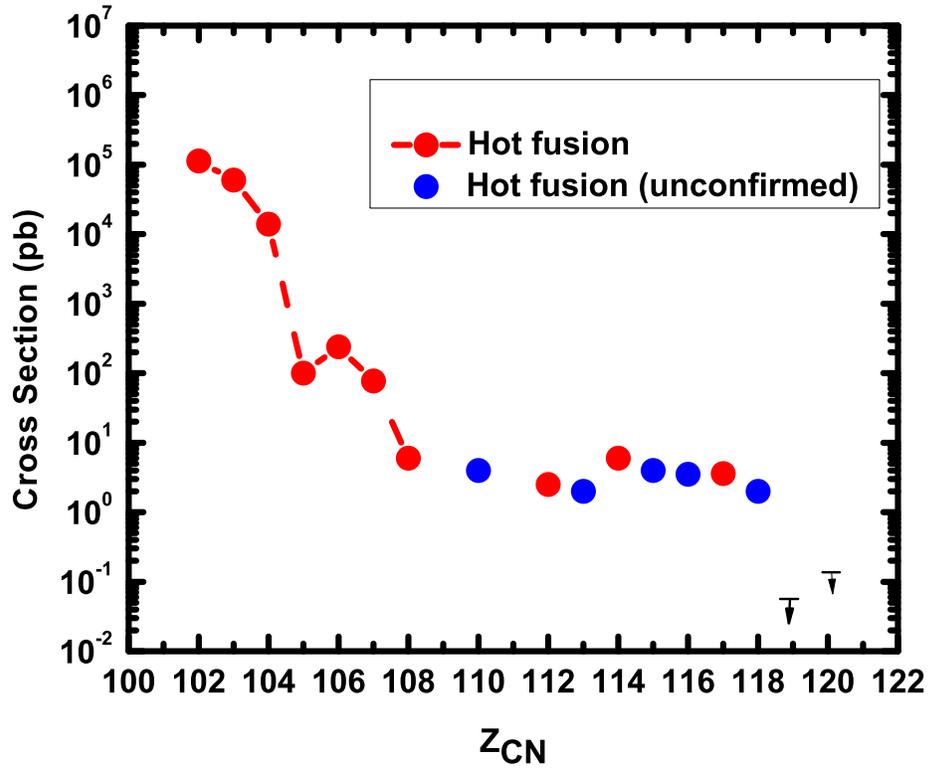}
\end{center}
\caption{(Color-online)Formation cross sections for new heavy elements
 using hot fusion reactions. }
\label{fig1}
\end{minipage}
\end{figure}

\begin{figure}[tbp]
\begin{minipage}{30pc}
\begin{center}
\includegraphics [width=150mm]{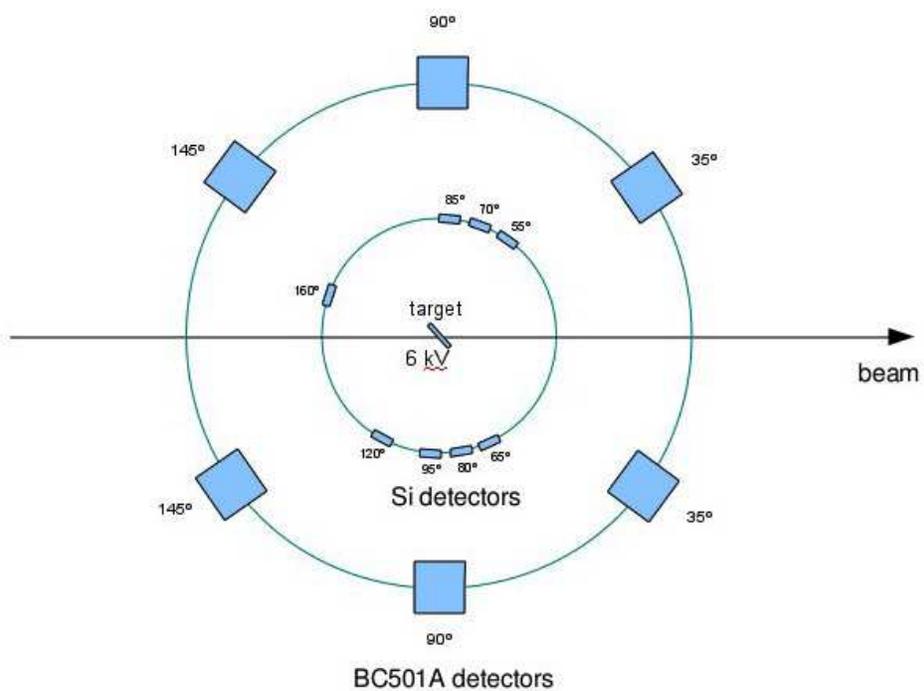}
\end{center}
\caption{(Color-online)Schematic diagram of the experimental apparatus. }
\label{fig2}
\end{minipage}
\end{figure}

\begin{figure}[tbp]
\begin{minipage}{30pc}
\begin{center}
\includegraphics [width=100mm]{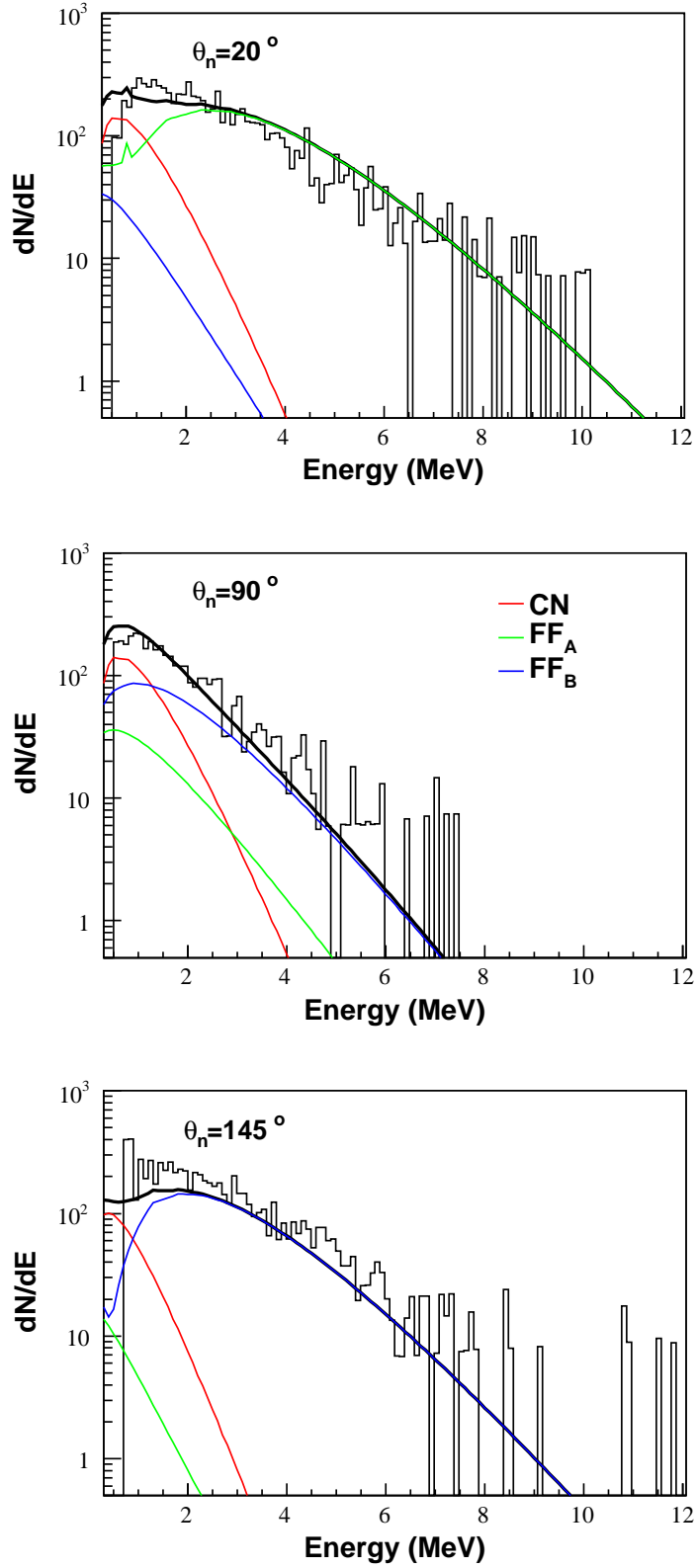}
\end{center}
\caption{(Color-online)The moving sources fits for various detector combinations for the $^{26}$Mg + $^{248}$Cm reaction.}
\label{fig3}
\end{minipage}
\end{figure}

\begin{figure}[tbp]
\begin{minipage}{30pc}
\begin{center}
\includegraphics [width=150mm]{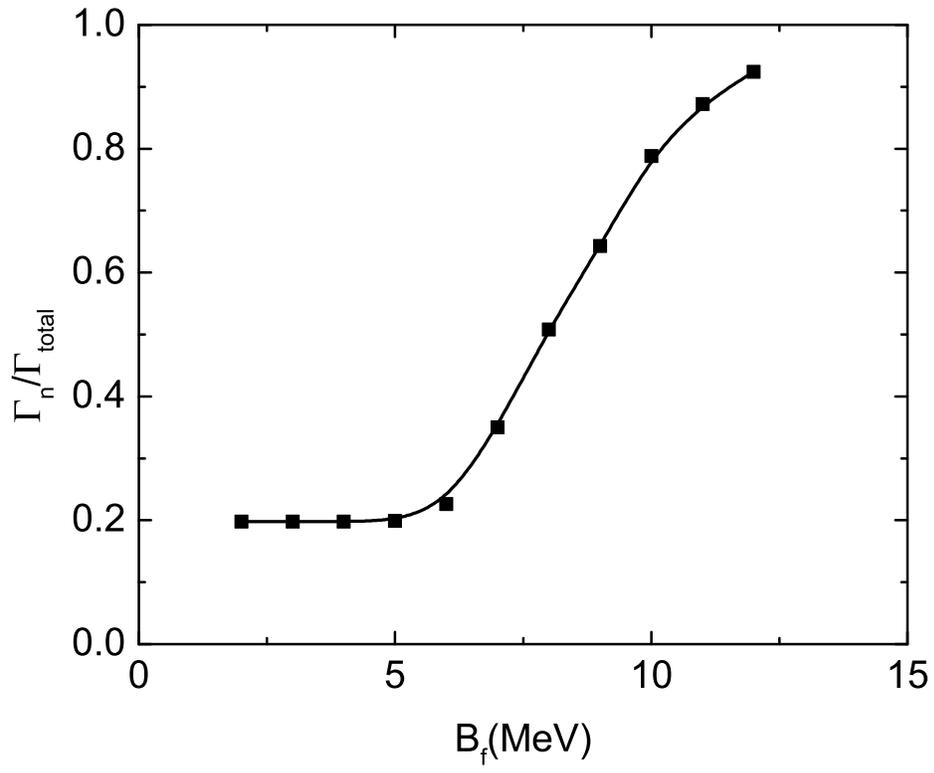}
\end{center}
\caption{The calculated variation of  $\Gamma$$_{n}$/$\Gamma$$_{total}$ with changes in B$_{f}$.}
\label{fig4}
\end{minipage}
\end{figure}

\end{document}